 \documentstyle[aps,epsf,eqsecnum,twocolumn]{revtex}

\begin{document}
\title{Localization of Eigenstates \& Mean Wehrl Entropy}
\author{Karol {\.Z}yczkowski$^{1,2}$}
\address{$^1$Instytut Fizyki im. Mariana Smoluchowskiego, \\
Uniwersytet Jagiello{\'n}ski, ul. Reymonta 4, 30-059 Krak{\'o}w, Poland}
\address{$^2$Centrum Fizyki Teoretycznej PAN \\
Al. Lotnik{\'o}w 32/44, 02-668 Warszawa, Poland}
\date{\today}
\maketitle

\begin{abstract}
Dynamics of a periodically time dependent quantum system is
reflected in
the features of the eigenstates of the Floquet operator. Of the special
importance are their localization properties quantitatively characterized by
the eigenvector entropy, the inverse
 participation ratio or the eigenvector
statistics. Since these quantities depend on the choice of the eigenbasis,
we suggest to use the overcomplete basis of coherent states, uniquely
determined by the classical phase space. In this way we define the mean
Wehrl entropy of eigenvectors of the Floquet operator and demonstrate that
this quantity is useful to describe quantum chaotic systems.
\end{abstract}


\vskip 0.4cm

\begin{center}
{\small e-mail: $^1$karol@tatry.if.uj.edu.pl}
\end{center}

\vskip 0.5cm


\section{Introduction}

Analysis of quantum chaotic systems is often based on the statistical
properties of the spectrum of the Hamiltonian $H$
 (in the case of autonomous
systems) or the Floquet operator $F$ (in the case of
 periodically perturbed
systems). In general, quantized analogous of classically chaotic systems
display spectral fluctuations conforming to the predictions of random
matrices. Depending on the geometrical properties of the system one uses
orthogonal, unitary or symplectic ensemble \cite{Ha91,CC96}.

Another line of research deals with eigenstates of the analyzed quantum
system. One is interested in their localization properties, which can be
characterized by the eigenvector distribution
\cite{KMH88,Bo91,Iz90,HZ90}
the entropic localization length \cite{CGIS90} or the inverse
participation
ratio \cite{He87}. All this quantities, however, are based on the
expansion
of an eigenstate in a given basis $\{\vec{b}_i\}$, which may be chosen
arbitrarily. If one chooses (with a bad will), $\{\vec{b}_i\}$ as the
eigenbasis
of $F$, all these quantities carry no information whatsoever. One may ask,
therefore, to what extend the quantitative analysis based on the
eigenvector statistics is reliable.

Let $G$ denote a unitary operator, such that $\{ \vec{b}_i\}$ is its
eigenbasis.
We showed \cite{KZ91,Z92,Zy93} that the eigenvector statistics of a
quantum map $F$ conforms to
 the prediction of random matrices, if operators $F$ and $G$
are {\sl relatively random}, i.e., their commutators are sufficiently
large.

In this paper we advocate an alternative method of solving the problems with
arbitrariness of the choice of the expansion basis. Instead of working in a
finite discrete basis, we shall use the coherent states expansion of the
eigenstates of $F$. For several examples of compact classical phase spaces
one may construct a canonical
 family of the generalized coherent states
\cite{Pe86}. Localization properties of any pure quantum state may be
characterized by the Wehrl entropy, equal to the average log of its overlap
with a coherent state \cite{We78,We91}. We propose to describe the structure
of a given Floquet operator $F$ by the mean Wehrl entropy of its
eigenstates. This quantity, explicitly defined without any
arbitrariness, is shown to be a useful indicator of quantum chaos.

This paper is organized as follows. In section II we review the definition
of the Husimi distribution, stellar representation, and the Wehrl entropy.
For concreteness we work with the $SU(2)$ vector coherent states, linked to
the algebra of the angular momentum operator and corresponding to the
classical phase space isomorphic with the sphere. In section III we
define the
mean Wehrl entropy of eigenstates and present analytical results obtained
for low dimensional Hilbert spaces. Exemplary application of this quantity
to the analysis of the quantum map describing the model of the periodically
kicked top is provided in section IV.

\section{Husimi distribution and stellar representation}

Consider a compact classical phase space $\Omega$, a classical area
preserving map $\Theta:\Omega \to \Omega$ and a corresponding quantum map $F$
acting in an $N$-dimensional Hilbert space ${\cal H}_N$. A link between
classical and quantum mechanics can be established via a family of
generalized coherent states $|\alpha \rangle$. For several examples of
the classical phase spaces there exist a canonical family of coherent
states.
It forms an overcomplete basis  and allows for an identity resolution
$\int_{\Omega} |\alpha \rangle \langle \alpha | d\alpha ={\bf 1}$. Any
mixed quantum state,
described by a density matrix $\rho$ can be represented by the
generalized Husimi distribution \cite{Hu40}, (Q-function)
\begin{equation}
H_{\rho}(\alpha ):= \langle \alpha |\rho |\alpha \rangle.  \label{hus2}
\end{equation}
The standard normalization of the coherent states,
$\langle \alpha | \alpha\rangle =1$,  assures that $\int_{\Omega}
H_{\rho}(\alpha)
d \alpha = 1$. For a pure quantum state $|\psi\rangle$ the Husimi
distribution is  equal
to the overlap with a coherent state $H_{\psi}(\alpha ):=
|\langle \psi |\alpha \rangle|^2$. Let us note
that the Husimi distribution was
successfully applied to study dynamical properties of quantized
chaotic systems \cite{Ta86,Zy87}.

Consider a discrete partition of the unity into $n$ terms; $\sum_{i=1}^n
p_i=1$. The Shannon entropy $S_d=-\sum_{i=1}^n p_i\ln p_i$ characterizes
uniformity of this partition. In an analogous way one defines the Werhl
entropy of a quantum state $\rho$ \cite{We78}
\begin{equation}
S_{\rho} = - \int_{\Omega} H_{\rho}(\alpha) \ln [H_{\rho}(\alpha)]
d\alpha,
\label{wer1}
\end{equation}
in which the summation is replaced by the integration over the classical
space $\Omega$. This quantity characterizes the localization properties of a
quantum state in the classical phase space. It is small for coherent states
localized in the classical phase space $\Omega$ and large for the
delocalized states. The maximal Wehrl entropy corresponds to the maximally
mixed state $\rho_*$, proportional to the identity matrix,
for which the Husimi distribution is uniform.

Although the notions of the generalized coherent states, the Husimi
distribution, and the Wehrl entropy are well defined for several
classical
compact phase spaces, in this work we analyze in detail only the most
important case $\Omega=S^2$. This phase space is typical to
physical problems involving spins, due to the algebraic properties of the
angular momentum operator $J$. In this case one uses the family of spin
coherent states $|\vartheta,\varphi \rangle $ localized at points
$(\vartheta,\varphi)$ of the sphere $S^2$. These states, also called
$SU(2)$
vector coherent states, were introduced by Radcliffe \cite{r71} and Arecchi
{\sl et al.} \cite{a72} and are an example of the general group theoretic
construction of Perelomov \cite{Pe86}.

Consider an $N=2j+1$ dimensional representation of the angular momentum
operator $J$. For a reference state one usually takes the maximal eigenstate
$|j,j\rangle$ of the component $J_z$. This state, pointing toward the "north
pole" of the sphere, enjoys the minimal uncertainty. The vector
coherent state represents the reference state rotated by the angles $%
\vartheta$ and $\varphi$. Its expansion in the basis $|j,m\rangle$, $%
m=-j,\dots,+j$ reads \cite{VS95}
\begin{eqnarray}
|\vartheta , \varphi \rangle =\sum_{m=-j}^{m=j}
\sin ^{j-m}({\frac \vartheta 2})\cos ^{j+m}({\frac \vartheta 2})
\times \nonumber \\
\exp \Bigl(i(j-m)\varphi  \Bigr)
\Bigl[ \Bigl({\ {\ {{ {2j  \atop j-m}
} } } }\Bigr)\Bigr]^{1/2}|j,m\rangle,  \label{thetrot}
\end{eqnarray}
where
$\int_0^{2 \pi} d\varphi
 \int_0^{\pi} \sin \vartheta d\vartheta
 |\vartheta, \varphi \rangle \langle \vartheta,\varphi |
(2j+1)/4\pi  ={\bf 1}$.

For the $SU(2)$ coherent state the distribution (\ref{hus2}) equals in this
case $H_{\psi}(\vartheta,\varphi):= |\langle \psi | \vartheta,
\varphi
\rangle|^2$. Two different spin coherent states overlap unless they are
directed into two antipodal points on the sphere.  The Husimi representation
of a spin coherent state has thus one zero (degenerated $N-1$ times)
localized at the antipodal point. In general,
 any pure quantum state can be
uniquely described by the set of $N-1$ points distributed over the sphere.
Some of these zeros may be degenerated, just as in the case of a coherent
state. This method of characterizing a pure quantum state is called the
stellar representation \cite{Penr,Leb}.

In the analyzed case of the classical phase space isomorphic with the
sphere $S^2$  the Wehrl
entropy (\ref{wer1}) of a state $\rho$ equals
\begin{equation}
S_{\rho} = - {\frac{2j +1 }{4 \pi}} \int_0^{\pi}
\! \! \sin \vartheta d\vartheta \!
\int_0^{2\pi} \! d\varphi H_{\rho}(\vartheta,\varphi) \ln \Bigl[
H_{\rho}(\vartheta,\varphi)\Bigr],  \label{wehr2}
\end{equation}
since the measure element $d \alpha$ is equal to
$(2j+1) \sin \vartheta d \vartheta d \varphi /4 \pi$.
Under this normalization the entropy of the maximally mixed state $\rho_*$
equals to $\ln N$.

The Husimi distribution of an eigenstate $|j,m\rangle $ may be computed
directly from the definition (\ref{thetrot}). Due to the rotational symmetry
it does not depend on the azimuthal angle $\varphi $
\begin{equation}
H_{|j,m\rangle }(\vartheta )= \sin^{2(j-m)}(\vartheta  / 2 )\cos
^{2(j+m)}( \vartheta / 2 )\Bigl({ {2j \atop {j-m}}
}\Bigr),  \label{husjz}
\end{equation}
which simplifies the computation of the Wehrl entropy. Simple integration
gives for the reference state  $|j,j\rangle $
\begin{equation}
S_{{\rm coh}}={\frac{N-1}{N}}={\frac{2j}{2j+1}}.  \label{lieb}
\end{equation}
Due to the rotational invariance the Wehrl entropy is the same for
any other coherent state.

\medskip

\begin{tabular}{|c|c|c|c|c|}
\hline
$N$ & $j$ & $m$ & $S_{|j,m\rangle }$ & ${\bar S}_{J_z}$ \\ \hline\hline
$2$ & $1/2$ & $1/2$ & $1/2$ & $1/2=0.5$ \\ \hline\hline
$3$ & $1$ & $1$ & $2/3$ & $1-\frac{\ln 2}{3} \approx 0.769$
 \\ \cline{3-4}
  &   & $0$ & $5/3-\ln 2$ &  \\ \hline\hline
$4$ & $3/2$ & $3/2$ & $3/4$ & $\frac{3}{2}-\frac{\ln 3}{2}
\approx 0.951$ \\ \cline{3-4}
   &    & $1/2$ & $9/4-\ln 3$ &  \\ \hline\hline
   & & $2$ & $4/5$ &  \\ \cline{3-4}
$5$& $2$ & $1$ & $79/30-\ln 4$ & $2-\frac{\ln 96}{5}
\approx 1.087 $ \\ \cline{3-4}
  &   & $0$ & $47/15-\ln 6$ &  \\ \hline\hline
 &  & $5/2$ & $5/6$ &  \\ \cline{3-4}
$6$& $5/2$ & $3/2$ & $35/12-\ln 5$ &
 $\frac{5}{2}-\frac{1}{3} \ln 50\approx 1.196$ \\
\cline{3-4}
 & & $1/2$ & $15/4-\ln 10$ &  \\ \hline
\end{tabular}

\smallskip

Table 1. Wehrl entropy $S_{|j,m\rangle}$ for the eigenstates of $J_z$
and its mean  ${\bar S}_{J_z}$
for $N=2,3,4,5,6$. Due to the geometrical symmetry
$S_{|j,m\rangle}=S_{|j,-m\rangle}$.

\medskip

The Wehrl entropies for other eigenstates of $J_{z}$ are collected in
Tab. 1
for some lower values of $N$. These results may be also obtained from
the general formulae provided by Lee \cite{Le88} for the Wehrl entropy
of the
pure states in the stellar representation. Eigenstate $|j,m\rangle $ is
represented by $j+m$ zeros at the south pole and $j-m$ zeros at the
north
poles. For $j=1/2$ ($N=2$) all the states are $SU(2)$ coherent, so their
entropies are equal. For $j=1$ ($N=3$) the coherent state $|1,1\rangle $ is
characterized by the smallest entropy, while the state $|1,0\rangle $ by the
largest (among the pure states).
 The larger $N$, the more place for a
various behaviour of pure states, measured by the values of $S$. The
axis of the Wehrl entropy is drawn schematically in Fig.1.

\vskip -1.5cm
\begin{figure}
\hspace*{-1.6cm}
\vspace*{-2.7cm}
\epsfxsize=9.5cm
\epsfbox{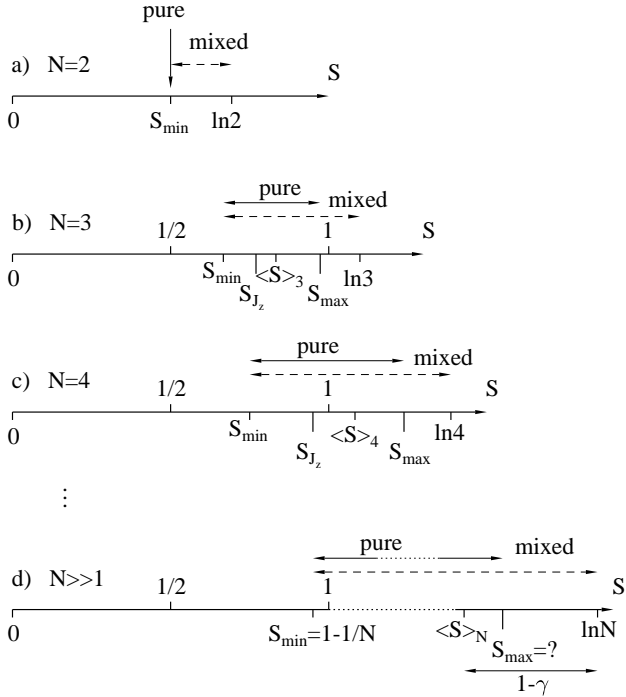} \\
\caption{Axis of Wehrl entropy for pure states
of $N$ dimensional Hilbert space;
a) $N=2$ for which
 ${\bar S}_{\rm min}={\bar S}_{\rm max}=1/2$;
 b) $N=3$ for which
${\bar S}_{\rm min}=2/3$,  $\bar{S}_{J_z}\approx 0.77$,
$\langle S \rangle_3 \approx 0.83$, $S_{\rm max}\approx
0.973$;
 c) $N=4$  for which
${\bar S}_{\rm min}=3/4$,  $\bar{S}_{J_z}\approx 0.95$,
$\langle S \rangle_4 \approx 1.08$, $S_{\rm max}\approx
1.24$;
  and d) $N>>1$, where  ${\bar S}_{\rm min}=1-1/N$
 while  $\langle S \rangle_N \approx \ln N -0.423$.
  }
\label{f1}
\end{figure}

It has been conjectured by Lieb \cite{Li78} that vector coherent
states are
characterized by the minimal value of the Wehrl entropy $S_{{\rm min}}=S_{%
{\rm coh}}$, the minimum taken over all mixed states.  For partial results
in the direction to prove this conjecture see
\cite{Scutar,Le88,Schupp}. It
was also conjectured \cite{Le88} that the states with possibly regular
distribution of all $N-1$ zeros of the Husimi function on the sphere
 are characterized by the
largest possible Wehrl entropy among all pure states $S_{\rm max}$.
Such a distribution of zeros is easy to specify for
$(N-1)=4,6,8,12,20$, which correspond to the Platonian polyhedra.
For $N=2$ all pure states are coherent, so $S_{\rm min}=S_{\rm
max}=1/2$. For $N=3$ the maximal Wehrl entropy
$S_{\rm max}=5/3-\ln 2\approx 0.97$ is achieved for the state
$|1,0\rangle$, for which the two zeros of Husimi function are localized
at the opposite poles of the sphere. For $N=4$ the state with three
zeros located at the equilateral
 triangle inscribed in a great circle is characterized by
$S_{\rm max}=21/8-2\ln 2\approx 1.24$. It will be interesting to
find
such maximally delocalized pure states for larger values of $N$, and to
study the dependence $S_{\rm max}$ of $N$.

Let us emphasize that for $N>>1$ the pure states exhibiting small Wehrl
entropy, (of the order of $S_{\rm min}$), are not typical. In the
stellar
representation coherent states correspond to coalescence of all $N-1$ zeros
of Husimi distribution in one point. In a typical situation the density
of the zeros is close to uniform on the sphere,
and the Wehrl entropy of such delocalized pure states is large.
A random state can be generated according to the natural uniform
masure on the space of pure
states by taking any vector of a $N \times N$  random matrix
distributed according to the Haar measure on $U(N)$.
 Averaging over this measure
 one may compute the mean Wehrl entropy $\langle
S\rangle_N $ of the pure states belonging to the $N$ dimensional
Hilbert space.
Such integration was performed in \cite{KMH88,J91,SZ98}
in a slightly different context leading to
\begin{equation}
\left\langle S\right\rangle_N=\Psi \left( N+1\right) -\Psi \left(
2\right) =\sum_{n=2}^{N}{\frac{1}{n}},  \label{wehmean}
\end{equation}
where $\Psi $ denotes the digamma function. Note that another normalization
of the coherent states used in Ref. \cite{SZ98}, leads to results shifted by
a constant $- \ln N$. Such a normalization
allows one for a direct comparison between the entropies
describing the states of various $N$.
 In the asymptotic limit $N\rightarrow \infty $ the mean
entropy $\langle S \rangle_N$
 behaves as $\ln N+\gamma -1\sim \ln N-0.42278$, which is close to
the maximal possible Wehrl entropy
 for mixed states $S_{\rho _{\ast }}=\ln N$. This result is
schematically marked in Fig 1d.

\section{Mean Wehrl entropy of eigenstates of quantum map}

Consider a quantum pure state in the $N$-dimensional Hilbert space. Its
Wehrl entropy computed in the vector coherent states representation may
vary  from $1-1/N$, for a coherent state, to the number of order of
$\ln N$,
for the typical delocalized state. This difference suggests a simple
measure
of localization of eigenstates of a quantum map $F$. Denoting its
eigenstates by $|\psi_i\rangle$; $i=1,\dots,N$ we define the mean Wehrl
entropy of eigenstates
\begin{equation}
{\bar{S}_F} ={\frac{1 }{N}} \sum_{i=1}^N S_{|\psi_i \rangle }.
\label{meawer}
\end{equation}
This quantity may be straightforwardly computed numerically for an
arbitrary quantum
map $F$. For quantum analogues of classically chaotic systems exhibiting no
time reversal symmetry all eigenstates are  delocalized. In this case the
mean Wehrl entropy of eigenvectors ${\bar S}_F$ fluctuates around
$\langle S \rangle_N \sim \ln N$.

In the opposite case of an integrable dynamics the eigenstates are, at least
partially, localized. A simple example is provided by any Hamiltonian
diagonal in the $J_{z}$ basis (or the basis of any other component of
$J$). The mean Werhl entropy
 $\bar{S}_{J_z}$ is given in table 1 for some values of $N$.
Further analysis shows \cite{SKZ99} that for larger $N$ the mean entropy
behaves as ${1 \over 2} \ln N$. This result has a simple
interpretation. Let us
divide the surface of the sphere into $N\sim \hbar ^{-1}$ cells. A typical
eigenstate of $J_{z}$ is localized in a longitudinal strip of a constant
polar angle $\theta $ and covers $\sqrt{N}$ of the cells, so its entropy
is of the order of $\ln \sqrt{N}$.

The quantity $\bar S$ is well--defined in a generic
case of operators $F$ with a nondegenerate spectrum.
In the case of the degeneracy, there
exists a freedom of choosing the eigenvectors; to cure this lack of
uniqueness we define ${\bar S}_F$ as the minimum over all possible sets
of eigenvectors of $F$.
Having a general definition of the mean Wehrl entropy of eigenvectors
of an arbitrary unitary operator, one
 may pose a question, for which operators $F_{\rm min}$
 ($F_{\rm max}$) of a fixed $N$ with a nondegenerate spectrum
 this quantity is the smallest
(the largest). It is clear that $\bar{S}_{F_{\rm min}}$ is larger than
$S_{\rm coh}$ (for $N>2$), since the set of any $N$ coherent states
 does not form an orthogonal basis. On the other hand, the
minimum is smaller
than $\bar{S}_{J_z}$,
as explicitly demonstrated in Appendix for $N=3$.
The value $\bar {S}_{F_{\rm max}}$ is larger than the average over the
random unitary matrices
$\langle S \rangle_{U(N)}=\langle S \rangle_N$ and smaller than
$S_{\rho_*}=\ln N$.

The mean Werhl entropy of the eigenstates ${\bar S}_F$
may be related
with the eigenvector statistics of the operator $F$. Let us expand a
given coherent state in the
eigenbasis of the Floquet operator,
 $|\alpha \rangle=\sum_{i=1}^N c_i(\alpha) |\psi_i\rangle $. The
dynamical properties of a quantum system are characterized locally \cite
{Zy90} by the Shannon entropy $S_s(\alpha):=-\sum_{i=1}^N
|c_i(\alpha)|^2 \ln  |c_i(\alpha)|^2$.
The mean Wehrl entropy may be thus written as an average over the phase
space
\begin{equation}
{\bar{S}_F} ={\frac{1 }{N}} \int_{\Omega} S_s(\alpha) d\alpha.
 \label{meawe2}
\end{equation}
This link is particularly useful to analyze the influence of the time
reversal symmetry. In presence of any generalized antiunitary symmetry
the operator $F$ may be described by the circular orthogonal ensemble
(COE).
There exists a symmetry line in the phase space and the coherent states
located along this line display  eigenvector statistics typical of COE
\cite{Zy91}. This symmetry is also visible in the stellar representation
of the eigenstates and manifests itself by a clustering of zeros
of Husimi functions \cite{BBL,BKZ97}. However, a typical coherent
state does not exhibit such a symmetry and its eigenvector statistics is
typical to the circular unitary ensemble
(CUE). Thus for a system with the time-reversal symmetry the mean Wehrl
entropy will be slightly smaller than for the analogous system with the
time reversal symmetry broken, but much larger than the Shannon entropy
of real eigenvectors of matrices pertaining to the orthogonal ensemble.

\section{Mean Wehrl entropy for the kicked top}

In order to demonstrate usefulness of the mean Wehrl entropy in the
analysis of quantum chaotic systems we present numerical results
obtained for the periodically kicked top. This model is very suitable
for investigation of quantum chaos \cite{HKS,Ha91}. Classical dynamics
takes place on the sphere, while the quantum map is defined in terms of
the components of the angular momentum operator $J$. The size of the
Hilbert space is determined by the quantum number $j$ and equals
$N=2j+1$.  One step evolution
operator reads $F_o=\exp(-ipJ_z)\exp(-ikJ_x^2/2j)$. For $p=1.7$ the
classical system becomes fully chaotic for the kicking strength
$k\approx 6$ \cite{HKS}. This system possesses a generalized antiunitary
symmetry and can be described by the orthogonal ensemble. The time
reversal symmetry may be broken by an additional kick \cite{Ha91}. The
system $F_u=F_o \exp(-ik'J_y^2/2j)$  pertains to CUE and
will be called the unitary top.

\vskip -0.2cm
\begin{figure}
\hspace*{-0.6cm}
\vspace*{0.4cm}
\epsfxsize=9.0cm
\epsfbox{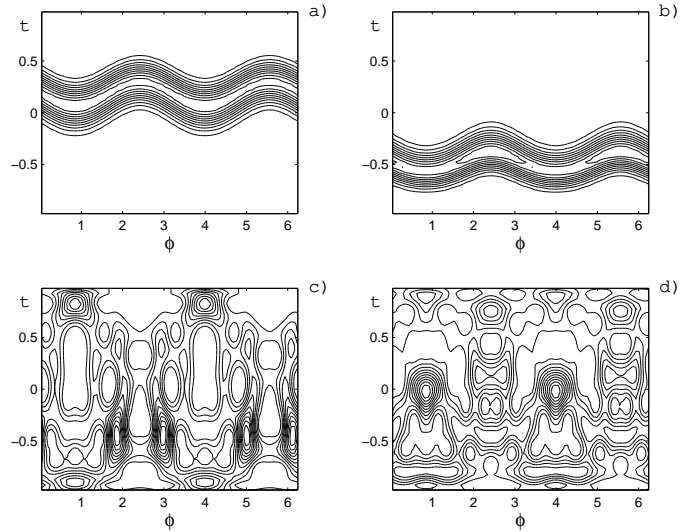} \\
\caption{Husimi distribution of exemplary eigenstates of the
Floquet
operator of the orthogonal kicked top for $N=62$ in the
dominantly regular
regime ($k=0.5$), a) and b), and chaotic regime ($k=8.0$),
c) and d).
The sphere is represented in a rectangular projection with
$t=\cos\vartheta$. }
\label{f2}
\end{figure}

Fig. 2 presents the Husimi distributions of two eigenstates of $F_o$
for $p=1.7$
in the regime of regular motion $(k=0.5)$ and two, for which
the classical dynamics is chaotic $(k=8.0$). In the quasiregular case
the eigenstates are localized close to parallel strips,
covered uniformly by the eigenstates of $J_z$. On the other
hand, the eigenstates of the chaotic map are delocalized at the entire
sphere. These differences are well characterized by the values of the
Wehrl entropies, equal correspondingly: a) $2.77$, b) $2.66$; and
c) $3.72$, d) $3.80$. The data, obtained for $N=62$, may be compared
with the mean entropy of the unperturbed system,
$\bar{S}_{J_z}\approx 2.465$,
the mean Wehrl entropy of chaotic system without time reversal symetry,
$\langle S \rangle_{62}\approx 3.712$, and the maximal entropy of
the mixed state, $S_{\rho_*}=\ln 62 \approx 4.1271$.

The above eigenstates are typical for both systems, and the other $60$
states display a similar character. The properties of all eigenstates
are thus described by the mean Wehrl entropy of eigenstates $\bar
{S}_F$. The dependence of this quantity on the kicking strength $k$ is
presented in Fig. 3. To show a relative difference between the entropies
typical to the regular dynamics we use the scaled coefficient
\begin{equation}
\mu (F) := { {\bar{S}}_F - {\bar{S}}_{J_z} \over
                \langle S \rangle_N - {\bar{S}}_{J_z} }.
\label{gamma}
\end{equation}
Per definition $\mu$ is equal to zero, if $F$ is diagonal in the
$J_z$ basis, which corresponds to the integrability. In the chaotic
regime $F$ is well described by CUE and $\mu$ is close to unity.
This is indeed the case for the unitary top with $k'=k/2$ and $k>6$.
The growth of $\mu$ is bounded and therefore it cannot, in general,
follow
the increase of the classical Kolmogorov--Sinai entropy $\Lambda$ (the
Lapunov exponent averaged over the phase space), which for the
classical system grows with the kicking strength $k$ \cite{Zy93}.
The data for the orthogonal top fluctuate below unity, due to existence
of the symmetry line. The difference between the coefficients $\mu$
obtained for both models does not depend on the kicking strength,
but decreases with $N$ and vanish in the classical limit $N\to \infty$.

\vskip -1.4cm
\begin{figure}
\hspace*{-1.6cm}
\vspace*{-6.6cm}
\epsfxsize=9.9cm
\epsfbox{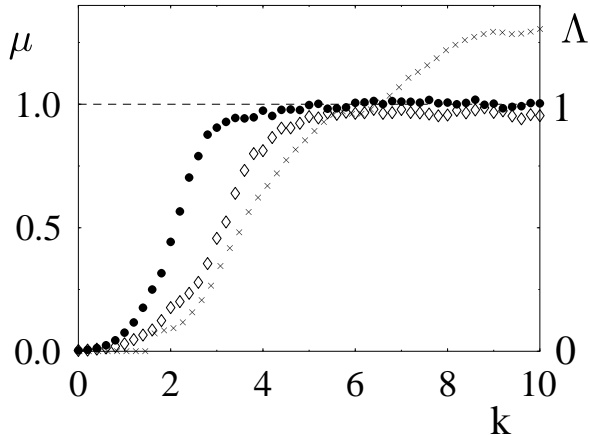} \\
\caption{Scaled mean Wehrl entropy $\mu$ of the eigenvectors
of the Floquet operator for the
kicked top as a function of the kicking strength $k$ for
$N=62$. The
data are obtained for two models: unitary top $(\bullet)$ and
orthogonal top $(\diamond)$.
The crosses denote values of the classical Kolmogorov-Sinai entropy
$\Lambda$, which characterize
the transition to chaos in the classical analogue of the
orthogonal top.
}
\label{f3}
\end{figure}

\section{Concluding remarks}

The Wehrl entropy of a given state characterizes its localization in the
classical phase space. We have shown that the mean Wehrl entropy
${\bar S}_F$ of eigenstates of a given evolution operator
$F$ may serve as a useful
indicator of quantum chaos. Let us emphasize that this quantity, linked
to the classical phase space by a family of coherent states, does not
depend on the choice of basis. This contrasts the others quantities,
like eigenvector statistics, localization entropy, inverse participation
ratio, often used to study the properties of eigenvectors.
It will be interesting to find the unitary operators (or rather the
repers) for which ${\bar S}_F$ is the smallest or the largest.

The mean Wehrl entropy of eigenstates enables one to detect the
transition from regular motion to chaotic dynamics. On the other hand, it
is not related to the classical Kolmogorov--Sinai entropy (or
to the Lapunov exponent), so it cannot be
used to measure the degree of chaos in quantum systems. Such a link with
the classical dynamics is established for
the {\sl coherent states dynamical entropy} of a given quantum map
\cite{SZ95,SZ98},
but this quantity is much more difficult to calculate.
Both these quantities characterize the
 {\sl global} dynamical properties of a quantum system, in contrast to
the entropy of Mirbach and Korsch \cite{MK95,MK98}, which describes the
{\sl local} features.

Mean Wehrl entropy  characterizes the structure of eigenvectors of $F$,
and is not related at all to the spectrum of this operator. Thus it is
possible to construct a unitary operator with a Poissonian spectrum and
the delocalized eigenvectors. Or conversely, one may find an operator
with spectrum typical to CUE and all eigenstates localized.
This shows that the relevant information concerning the dynamical
properties of a quantum system described by an unitary evolution
operator $F$ is contained as well in its spectrum and in its eigenstates.

 I am indebted to W. S{\l}omczy{\'n}ski for
fruitful discussions and a constant interest in the progress of this
research. I am also thankful to M. Ku{\' s} and P. Pako{\' n}ski for
helpful remarks.
It is a pleasure to thank Bernhard Mehlig for the invitation to Dresden
and the Center for Complex Systems for a support during the workshop.
Financial support from Polski Komitet Bada{\'n}
Naukowych in Warsaw under the grant no 2P-03B/00915
is gratefully acknowledged.

\appendix

\section{Minimal mean Werhl entropy for $N=3$}

In the case $j=1$ ($N=3$) any pure state may be described by the
position of two points on the sphere - zeros of the
corresponding Husimi distribution. In the stellar representation the
eigenstates $|m\rangle$ of $J_z$ are described by $1+m$ zeros at the
south pole of the sphere, and $1-m$ zeros at the north pole, with
$m=-1,0,1$.
 The mean Werhl entropy of eigenstates of $J_z$ is
equal to $1-(\ln 2)/3\approx 0.769$. We find another set of three
orthogonal pure states, characterized by a smaller value of $\bar{S}$,
by allowing to move one of the  zeros of the Husimi representation
along the meridian $\varphi=0$ (and $\varphi=\pi$).
 In other words, let us define two orthogonal states
\begin{eqnarray}
|\chi_{+}\rangle:= \cos\chi |1\rangle +  \sin\chi |0\rangle,
 \nonumber \\
|\chi_{-}\rangle:= -\sin\chi |1\rangle +  \cos\chi |0\rangle.
\label{alpha}
\end{eqnarray}
Keeping  the state $|-1\rangle$ fixed we define thus
a one parameter family of orthogonal basis $O_{\chi}=\{|-1\rangle,
|\chi_-\rangle, |\chi_+\rangle\}$.
The Husimi distributions of the state $|\chi_{-} \rangle$
has two zeros at the polar angles
$\{ \theta_-,\pi\}$, while the zeros of $|\chi_+ \rangle$
are located at  $\{ \pi, 2\pi-\theta_+\}$.
 Here  $\theta_-=(2-c^2)/(2+c^2)$ and
 $\theta_+=(2c^2-1)/(2c^2+1)$ with $c=\tan\chi$.
The angle $\theta$
 belongs to $[0,2\pi)$ and the $\theta=\pi$ represents the
south pole, while $\theta=0$ (or $\theta=2\pi$) denotes the north pole.
The location of zeros of Husimi representation of eigenstates of $J_z$
and $O_{\pi/4}$ is presented in Fig.4.

\vskip -1.4cm
\begin{figure}
\hspace*{-1.9cm}
\vspace*{-6.6cm}
\epsfxsize=9.9cm
\epsfbox{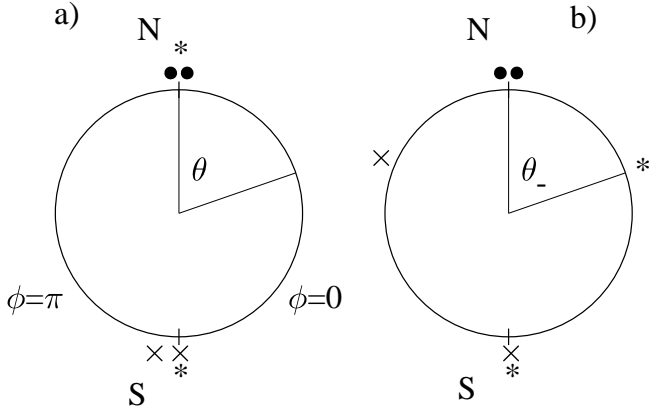} \\
\caption{Cross-section of the Bloch sphere along the meridian
$\phi=0$ $(\phi=\pi)$. a) zeros of the eigenstates of the operator $J_z$:
 $|-1\rangle$ $(\bullet)$, $|0\rangle$ $(*)$, and
$|1\rangle$ $(\times)$; b) zeros of the eigenstates of $O_{\pi/4}$:
$|-1\rangle$ $(\bullet)$, $|\chi_-\rangle$ $(*)$, and
 $|\chi_+\rangle$ $(\times)$.
}
\label{f4}
\end{figure}

In the paper of Lee \cite{Le88} one finds the Wehrl entropy of a $N=3$
state expressed as a function of the angle $\omega$ between both zeros
of the Husimi distribution:
\begin{equation}
S={{2/ 3} +{\sigma / 6} \over  1 -{\sigma / 2} }
  + \ln \Bigl( 1 -{\sigma \over 2} \Bigr),
\label{lee3}
\end{equation}
where $\sigma=\sin^2(\omega/2)$.
The equivalent result was earlier established by Scutaru \cite{Scutar},
$S=2/3+[a-\ln(1+a)]$, where  $a =\sigma/(2-\sigma)$.
Substituting for $\omega$ the angles $\pi-\theta_{\pm}$ we get explicit
formulae for the Wehrl entropy of each state
$|\chi_{\pm}\rangle$, which
allow us to write the mean entropy
${\bar{S}}_{O_{\chi}}$ as a function of the parameter $\chi$.
 The minimum
is achieved for $\chi=\pi/4$ (so $c=1$ and
 $\cos\theta_-\cos\theta_+ =1/3$) and reads
${\bar{S}}_{O_{\pi/4}}=1-[\ln(9/4)]/3 \approx0.730$.
This results shows that the eigenbasis of  $J_z$ does not provide the
reper characterized by the minimal mean Wehrl entropy,
but it does not solve the problem of finding the global minimum.
One can expect, that this minimum will occur for a set of $N$
mutually orthogonal states, each of them
as close to the coherent state, as possible.

\end{document}